\documentstyle[aps,preprint]{revtex}

\begin{document}

\title{Wave function of string and membrane and spacetime geometry}

\author{Hongwei Xiong, Shujuan Liu}
\address{Department of Applied Physics, Zhejiang University of Technology, Hangzhou, 310032, China}

\date{\today}

\maketitle

\begin{abstract}
A first-quantized string (and membrane) theory is developed here by using a
general wave function of the string (and membrane), analogously to the
first-quantized quantum theory of a point particle. From the general wave
function of the string (and membrane), the properties of the string (and
membrane) such as its relation to Bosons, Fermions and spacetime are
investigated. The string and membrane wave functions are found to be very
useful and we can deduce Klein-Gordon equation, Dirac equation and the
fundamental property of the spacetime from this new starting point.
\end{abstract}

{PACS number(s): 03.65.-w, 11.25.-w, 03.65.Pm}


\narrowtext

\newpage

During the last twenty years, we witnessed the rapid development \cite
{GREEN,POL} of superstring theory which is trying to become a theory of
everything. Due to the first and second revolutions of the superstring,
although a lot of miraculous characteristics of the superstring such as
string duality \cite{HULL,TOWN,WITTEN} are found, we still do not know the
first law of the superstring. To describe the string based on the general
quantum mechanical postulates, there are two approaches to investigate the
string theory. One is the first-quantized string theory where the string is
firstly regarded as a classical object, and the quantum property of the
string is obtained by using the quantization method. The other is the
second-quantized string field theory where the dynamical variables are
regarded as the functional of the string coordinates. In the present work,
we try to develop a first-quantized string (and membrane) theory by
proposing a general wave function of the string (and membrane), analogously
to the wave function of the point particle. The string (and membrane) is
regarded as an inner structure of the membrane (and string) wave function,
and is found to be relevant closely to the properties of bosons, fermions
and spacetime. Through the investigation of the differential equation of the
wave functions of membranes and strings, the Klein-Gordon equation for
bosons and Dirac equation for fermions are deduced in a nature way.

It is well known that the wave function of a point particle is $\Psi \left(
X^{\mu }\right) $ ($\mu =1,\cdots ,N$ for a $N-$dimensional spacetime.).
This sort of wave function is generalized to the case of the multi-component
wave function $\Psi _{i}\left( X^{\mu }\right) $ ($i=1,\cdots ,M$) which
describes the particles with nonzero spin. In this paper, the wave function
description of the point particle is generalized to develop a general
membrane (and string) theory. The wave function of the membrane is supposed
here to be $\Psi _{i}\left( X^{\mu }\left( \tau ,\sigma _{1},\cdots ,\sigma
_{l}\right) \right) $, and this would be the starting point to describe both
bosons and fermions. Here $\tau $ and $\sigma _{1},\cdots ,\sigma _{l}$ are
real parameters to show the inner freedom of the membrane. As a special
case, the string can be investigated when two parameters $\tau $ and $\sigma 
$ are used. Note that, when this sort of wave function of the membrane is
used, the membrane is no more regarded as an entity of the ordinary membrane
theory. In the ordinary first-quantized membrane theory, the membrane is
firstly regarded as a classical oscillating membrane, and the method of
quantization is used to reveal the quantum character of the membrane. In
this paper, we will try to reveal the character of the membrane and its
relation to the spacetime directly from the wave function of the membrane.
This can be regarded as a postulate in the present theory.

\vspace{1pt}Firstly, we will investigate the quantum properties of strings
from the string wave function. Suppose $\Psi _{M}^{T}=(\Psi _{1}\left(
X^{\mu }\left( \tau ,\sigma \right) \right) ,\cdots ,\Psi _{M}\left( X^{\mu
}\left( \tau ,\sigma \right) \right) )$. Here $T$ is a transpose of
matrixes. For this sort of string wave function, we consider here the
following general linear second-order partial differential equation:

\begin{equation}
\Xi _{1}\frac{\partial ^{2}\Psi _{M}\left( X^{\mu }\left( \tau ,\sigma
\right) \right) }{\partial \tau ^{2}}+\Xi _{2}\frac{\partial ^{2}\Psi
_{M}\left( X^{\mu }\left( \tau ,\sigma \right) \right) }{\partial \sigma ^{2}%
}=m^{2}\Psi _{M}\left( X^{\mu }\left( \tau ,\sigma \right) \right) ,
\label{fundamental}
\end{equation}
where $\Xi _{1}$ and $\Xi _{2}$ are $M\times M$ matrixes. After a simple
calculation, the above equation takes the form

\begin{equation}
\Xi _{1}\frac{\partial ^{2}\Psi _{M}}{\partial X^{\mu }\partial X^{\nu }}%
\frac{\partial X^{\mu }}{\partial \tau }\frac{\partial X^{v}}{\partial \tau }%
+\Xi _{2}\frac{\partial ^{2}\Psi _{M}}{\partial X^{\mu }\partial X^{\nu }}%
\frac{\partial X^{\mu }}{\partial \sigma }\frac{\partial X^{v}}{\partial
\sigma }+\Xi _{1}\frac{\partial \Psi _{M}}{\partial X^{\mu }}\frac{\partial
^{2}X^{\mu }}{\partial \tau ^{2}}+\Xi _{2}\frac{\partial \Psi _{M}}{\partial
X^{\mu }}\frac{\partial ^{2}X^{\mu }}{\partial \sigma ^{2}}=m^{2}\Psi _{M}.
\label{partical2}
\end{equation}
From the above equation, we have

\begin{equation}
\Xi _{1}\frac{\partial \Psi _{M}}{\partial X^{\mu }}\frac{\partial
^{2}X^{\mu }}{\partial \tau ^{2}}+\Xi _{2}\frac{\partial \Psi _{M}}{\partial
X^{\mu }}\frac{\partial ^{2}X^{\mu }}{\partial \sigma ^{2}}=0,  \label{main1}
\end{equation}
and

\begin{equation}
\Xi _{1}\frac{\partial ^{2}\Psi _{M}}{\partial X^{\mu }\partial X^{\nu }}%
\frac{\partial X^{\mu }}{\partial \tau }\frac{\partial X^{v}}{\partial \tau }%
+\Xi _{2}\frac{\partial ^{2}\Psi _{M}}{\partial X^{\mu }\partial X^{\nu }}%
\frac{\partial X^{\mu }}{\partial \sigma }\frac{\partial X^{v}}{\partial
\sigma }=m^{2}\Psi _{M}.  \label{main2}
\end{equation}

To have a nontrivial solution of $\Psi _{M}$ and $X^{\mu }\left( \tau
,\sigma \right) $, from Eq. (\ref{main1}), we have $\Xi _{1}=-\Xi _{2}$ and
get the following equation about strings:

\begin{equation}
\partial ^{2}X^{\mu }/\partial \tau ^{2}-\partial ^{2}X^{\mu }/\partial
\sigma ^{2}=0.  \label{string}
\end{equation}
It is not difficult to confirm that there is not a nontrivial solution of
strings if $\Xi _{1}=\Xi _{2}$. We see that the equation of the string is a
nature result when the string wave function is used to describe the
microscopic world. In this situation, the parameters $\tau $ and $\sigma $
of the string wave function are symmetry breaking. This implies the symmetry
breaking of the spacetime, i.e., for a $N-$dimensional spacetime, there
should be a one-dimensional time and $\left( N-1\right) -$dimensional space.

For Eq. (\ref{string}), at first, we consider the open string with Neumann
boundary condition $X^{\prime \mu }\left( \tau ,\sigma \right) |_{\sigma
=0,\pi }=0$. Under this boundary condition, the solution becomes:

\begin{equation}
X^{\mu }\left( \tau ,\sigma \right) =\xi _{0}^{\mu }+\xi _{1}^{\mu }\tau
+\sum_{n=1}^{\infty }\left( \eta _{n_{1}}^{\mu }\cos n\tau +\eta
_{n_{2}}^{\mu }\sin n\tau \right) \cos n\sigma .  \label{solution1}
\end{equation}
Note that for a classical string, the parameters $\xi _{0}^{\mu }$, $\xi
_{1}^{\mu }$, $\eta _{n_{1}}^{\mu }$, $\eta _{n_{2}}^{\mu }$ are determined
by the initial conditions ($X^{\mu }\left( \tau =0,\sigma \right) $ and $%
X^{\prime \mu }\left( \tau =0,\sigma \right) |_{\sigma }$) of the
oscillating string. In the case of quantum string introduced here, however,
there would be a much strong confinement condition on these parameters
through the investigation of Eq. (\ref{partical2}), and this confinement
would relevant closely to the properties of the spacetime.

For the oscillating mode $n$ of the open string, $X_{n}^{\mu }\left( \tau
,\sigma \right) =\left( \eta _{n_{1}}^{\mu }\cos n\tau +\eta _{n_{2}}^{\mu
}\sin n\tau \right) \cos n\sigma $. Substituting it into Eq. (\ref{main2}),
we have

\begin{equation}
\Xi _{1}\frac{\partial ^{2}\Psi _{M}}{\partial X^{\mu }\partial X^{\nu }}%
\chi _{n}^{\mu \upsilon }=m^{2}\Psi _{M},  \label{condi1}
\end{equation}
where

\[
{\ \chi _{n}^{\mu \upsilon }=\left( \eta _{n_{1}}^{\mu }\sin n\tau -\eta
_{n_{2}}^{\mu }\cos n\tau \right) \left( \eta _{n_{1}}^{v}\sin n\tau -\eta
_{n_{2}}^{v}\cos n\tau \right) n^{2}\cos ^{2}n\sigma - } 
\]

\begin{equation}
\left( \eta _{n_{1}}^{\mu }\cos n\tau +\eta _{n_{2}}^{\mu }\sin n\tau
\right) \left( \eta _{n_{1}}^{v}\cos n\tau +\eta _{n_{2}}^{v}\sin n\tau
\right) n^{2}\sin ^{2}n\sigma .  \label{condi2}
\end{equation}
We see that for $\Psi _{M}$ having a nontrivial solution, $m$ must be zero
and $\eta _{n_{1}}^{\mu }=\eta _{n_{2}}^{\mu }(=\eta _{n}^{\mu })$ (or $\eta
_{n_{1}}^{\mu },\eta _{n_{2}}^{\mu }=0$). In this case, for the oscillating
mode $n$, every component of $\Psi _{M}$ should satisfy the following
equation:

\begin{equation}
\frac{\partial ^{2}\Psi _{i}}{\partial X^{\mu }\partial X^{\nu }}\Gamma
_{n}^{\mu \upsilon }=0,  \label{klein}
\end{equation}
where $\Gamma _{n}^{\mu \upsilon }=\eta _{n}^{\mu }\eta _{n}^{v}$.

Eq. (\ref{klein}) is the differential equation of $\Psi _{i}$ about the
spacetime coordinates $X^{\mu }$. In a sense, in this equation, the inner
freedom of the string is omitted for the time being. However, the inner
motion of the string will give a strong confinement on the above equation
and even the properties of the spacetime. There are several results we can
deduce from the above equation: (i) $\Gamma _{n}^{\mu \upsilon }$ is a
second-rank symmetric tensor. For this sort of symmetric tensor, after a
matrix transformation, it becomes a diagonal matrix which can be regarded as
exactly the metric tensor $g^{\mu \nu }$ ($g^{\mu \nu }=U^{T}\Gamma
_{n}^{\mu \upsilon }U$ with $U$ a transformation matrix) of the spacetime.
We see that the geometry of the spacetime is determined by the oscillating
properties of the string due to the fact that the magnitude of the
oscillating string determines the metric tensor of the spacetime. The
invariance of Eq. (\ref{klein}) under Lorentz transformation is very obvious
based on the analysis here. (ii) The mass of this sort of string is zero,
even for any oscillating mode of the string. Thus, Eq. (\ref{klein}) is
exactly the well-known Klein-Gordon equation describing massless bosons.
(iii) From Eqs. (\ref{main2}) and (\ref{solution1}), when various
oscillating modes coexist, it is not difficult to confirm that the magnitude
of various oscillating modes must be identical and therefore correspond to
identical spacetime geometry. If various oscillating modes of the string are
regarded as various type of particles, the researches here show that various
type of particles must experience identical spacetime geometry. A character
obtained directly is the quantization of the magnitude of various
oscillating modes. Another implication would be the pith of the general
relativity, i.e., for any type of particles, its motion can be described by
the pure geometry of the spacetime.

In the case of a closed string with the periodicity condition $X^{\mu
}\left( \tau ,\sigma \right) =X^{\mu }\left( \tau ,\sigma =\pi \right) $, it
is straightforward to confirm that the discussions and results of the open
string hold also for the closed string. For both open and closed strings,
when the second-order partial differential equation of the string wave
function is used, the spin would be an integer number by calculating the
angular momentum of the string, i.e., $\Psi _{M}$ given by Eq. (\ref
{fundamental}) describes massless bosons. In addition, these discussions and
results hold too when the general membrane wave function is investigated.

Now we turn to investigate the first-order partial differential equation of $%
\Psi _{M}$. In this situation, we have

\begin{equation}
\Lambda _{1}\frac{\partial \Psi _{M}}{\partial X^{\mu }}\frac{\partial
X^{\mu }}{\partial \tau }+\Lambda _{2}\frac{\partial \Psi _{M}}{\partial
X^{\mu }}\frac{\partial X^{\mu }}{\partial \sigma }=m\Psi _{M}.
\label{fermion1}
\end{equation}
To have a solution of $\Psi _{M}$, $X^{\mu }$ should satisfy $\partial
X^{\mu }/\partial \tau =const$ and $\partial X^{\mu }/\partial \sigma =const$%
. In this situation, if the boundary condition of the open or closed strings
is used, we have

\begin{equation}
X^{\mu }\left( \tau ,\sigma \right) =\xi _{0}^{\mu }+\xi _{1}^{\mu }\tau .
\label{fermi2}
\end{equation}
From the above equation, Eq. (\ref{fermion1}) takes the form

\begin{equation}
\Lambda ^{\mu }\frac{\partial \Psi _{M}}{\partial X^{\mu }}=m\Psi _{M},
\label{fermion3}
\end{equation}
where $\Lambda ^{\mu }=\Lambda _{1}\xi _{1}^{\mu }$.

From the form of $\Lambda ^{\mu }$, we can see clearly the covariant
property of the matrixes $\Lambda ^{\mu }$. We know Eq. (\ref{fermion3}) is
the well-known Dirac equation which is used to describe the fermions with
nonzero mass. Note that in the investigation here about fermions, there is
in fact a wrong discussion. The matrix in Eq. (\ref{fermion3}) is $\Lambda
^{\mu }=\Lambda _{1}\xi _{1}^{\mu }$. It is obvious that $\Lambda ^{\mu }$
has only one linear independent matrix, while in the Dirac equation, four
linear independent matrixes are needed to show the spin freedom of fermions.
To generate four linear independent matrixes, the membrane wave function
should be $\Psi _{M}^{T}=(\Psi _{1}\left( X^{\mu }\left( \tau ,\sigma
_{1},\sigma _{2},\sigma _{3}\right) \right) ,\cdots ,\Psi _{M}\left( X^{\mu
}\left( \tau ,\sigma _{1},\sigma _{2},\sigma _{3}\right) \right) )$. In this
situation, $X^{\mu }\left( \tau ,\sigma _{1},\sigma _{2},\sigma _{3}\right)
=\xi _{0}^{\mu }+\xi _{1}^{\mu }\tau +\xi _{2}^{\mu }\sigma _{1}+\xi
_{3}^{\mu }\sigma _{2}+\xi _{4}^{\mu }\sigma _{3}$ and $\Lambda ^{\mu
}=\Lambda _{1}\xi _{1}^{\mu }+\Lambda _{2}\xi _{2}^{\mu }+$ $\Lambda _{3}\xi
_{3}^{\mu }+\Lambda _{4}\xi _{4}^{\mu }$. For this sort of three-dimensional
membrane to exist in the world we live, the dimensionality of the spacetime
must be larger than or equal to four (three-dimensional space and
one-dimensional time). Note that when the case of fermions is discussed, the
nontrivial solution is a nonoscillating string. From the result given by $%
X^{\mu }\left( \tau ,\sigma _{1},\sigma _{2},\sigma _{3}\right) $, there is
not a boundary condition for the membrane. In fact, the size of the membrane
can be finite or even infinite. This will not lead to conceptual difficulty
when the membrane is regarded as an inner structure of the membrane wave
function. We know for the wave function of a point particle, the wave
function can be nonzero in the whole world and this does not lead to any
problem. The membrane of the fermions discussed here is a nonoscillating
one. Maybe this would make superstring theorists a little distressed.
Nevertheless, the membrane can still oscillate in the extra warped space,
but not in the macroscopic four-dimensional spacetime. We see that, on the
one hand the finite size of the membrane is necessary to generate the Dirac
equation, and on the other hand this unique characteristic can be used
exactly to overcome the well-known ultraviolet divergences .

\vspace{1pt}In conclusion, the membrane wave function is proposed and used
to give an unified description of bosons and fermions. The general wave
function of the membrane (and string) proposed here is quite different from
the wave functional description of the membrane in the ordinary string field
theory, and can be regarded as a first-quantized string theory. Starting
from the membrane wave function, the fundamental property of the spacetime
can be investigated in a concise way. These results show that membrane (and
string) is necessary to describe the world we live, even when the problem of
the quantum gravity is not considered. Of course we do not propose a theory
of everything, what we try to argue here is that it is very useful to regard
the membrane as an inner structure of the general membrane wave function,
and expect that this new starting point would be helpful to answer the
question what the membrane is. Obviously, the strength of the superstring
and beautiful mathematics of the warped high-dimensional spacetime developed
in the last twenty years would still play an important role in the theory
proposed here. In addition, by investigating the general membrane wave
function, it would be an interesting future work to reconsider the
quantum-mechanical postulates.


\section*{Acknowledgments}


This work was supported by Natural Science Foundation of China under grant
Number 10205011.

\end{document}